\documentclass{elsart}

\usepackage{amsmath,amssymb,graphicx,color}

\def\cf{cf.}
\def\eg{e.g.}  
\def\eq#1{(Eq. \ref{#1})} 
 
\def\fig#1{Fig. \ref{#1}} 
\def\sec#1{Sec. (\ref{#1})}

\def\registered{{\ooalign{\hfil\raise .00ex\hbox{\scriptsize R}\hfil\crcr\mathhexbox20D}} }

\newcommand{\Czero}{\set{C}^0}                  
\newcommand{\grad}{\vec{\nabla}}                
\newcommand{\Hone}{\set{H}^1}                   
\newcommand{\set}[1]{\mathbb{#1}}               

\begin{document}

\begin{frontmatter}


\title{High-order low-storage explicit Runge-Kutta 
schemes for equations with quadratic nonlinearities}


\author[N,E]{Marc -E. Brachet}
\author[N,D]{Pablo D. Mininni}
\author[N]{Duane Rosenberg}
\author[N]{Annick Pouquet}. 

\address[N]{TNT/IMAGe, National Center for Atmospheric Research, P.O. Box 3000, Boulder, 
CO 80307-3000, USA}
\address[E]{Laboratoire de Physique Statistique de l'\'Ecole Normale Sup\`erieure, 
associ\'e au CNRS et aux Universit\'es ParisVI et VII, 24 Rue Lhomond, 75231 Paris, France}
\address[D]{Departamento de F\'\i sica, Facultad de Ciencias Exactas y Naturales, Universidad de Buenos Aires, Ciudad Universitaria, 1428 Buenos Aires, Argentina}
\begin{abstract}
We show in this paper that third- and fourth-order low storage Runge-Kutta algorithms can be built specifically for quadratic nonlinear operators, at the expense of roughly doubling the time needed for evaluating the temporal derivatives. The resulting algorithms are especially well suited for computational fluid dynamics. Examples are given for the H\'enon-Heiles Hamiltonian system and, in one and two space dimensions, for the Burgers equation using both a pseudo-spectral code and a spectral element code, respectively. The scheme is also shown to be practical in three space solving the incompressible Euler equation using a fully parallelized pseudo-spectral code.
\end{abstract}

\end{frontmatter}

\section{Introduction}
The explicit Runge--Kutta (ERK) method has a long and illustrious history in computational
science and engineering. For example, these schemes are often selected in high spatial resolution studies
of turbulence, in which the explicit nature of the scheme is used specifically
in order to capture all time scales in the manner of direct numerical simulation (DNS).
In the petascale era extremely high spatial resolutions can be reached, and 
improved accuracy in the time integration will also become necessary. 
The National Science Foundation has issued a baseline capability for its peta-scale initiative
whereby a computation of homogeneous turbulence on a uniform grid of $12288^3$ points and at a Taylor 
Reynolds number $R_{\lambda}\sim 2000$ should be accomplished in forty wall clock hours. In view of the 
large arrays to be stored for such a computation, and in view of the large number of operations 
required in order to arrive at a time of order unity required for this initiative, it is clear 
that an accurate, low--storage and reasonably simple scheme is imperative if this challenge is to be met.

While there are many references to Runge-Kutta schemes in the scientific and engineering 
literature, we mention several that are of particular interest in this paper.
It is well known that the standard fourth-order Runge-Kutta method can be evaluated
using three levels of storage \cite{blum}, and several low
storage versions of Runge-Kutta methods are considered in \cite{wili}. 
In reference \cite{Calvo}, new low storage methods adapted to acoustic problems is presented.
Obviously, the stability of the integration schemes is an important issue.
For high--order pseudo--spectral advection type problems this topic is also examined in detail 
in \cite{Renaut}; we consider here the problem of the stability of the algorithms empirically rather than from an analytical perspective.

In this work we begin with an algorithm (see Eq. \ref{eq_JSTLOOP} in \sec{ssec:jstloop}) which is  attributed to Jameson, Schmidt and Turkel \cite{JST} in the text of \cite{Canuto-Hussaini}. We do not find this algorithm in \cite{JST}, and are thus uncertain why the latter text would make this attribution. Nevertheless, we refer in the following to the algorithm given by Eq. (\ref{eq_JSTLOOP}) as the JST algorithm. Ostensibly, the JST algorithm requires only two levels of storage, but is of arbitrary order only for time--dependent linear problems. We show here that when the right hand side is nonlinear, corrections to the JST algorithm are needed if one wants to go beyond second order. 
The purpose of the present note is, after showing the limitations of the original algorithm, to compute the first corrections to the algorithm for quadratic nonlinearities as encountered for example in the modeling of incompressible flows. Our main result is that they can be implemented while preserving the low storage requirements. 

The paper is organized as follows. Section \ref{SEC:GenForm} contains the general formulation of our 
new time-stepping algorithm. Numerical applications to conservative and dissipative systems are 
provided in Section \ref{SEC:NumRes}.  Finally,  in Section \ref{SEC:Conc} we present our conclusion.

\section {General formulation} \label{SEC:GenForm}

\subsection {Basic definitions}
Our starting point will be the following equation of motion
for the vector ${\bf u}$
\begin{equation}
{d{\bf u}\over {dt}}={\bf F}({\bf u}),
\label{eq_evol}
\end{equation}
with
\begin{equation}
{\bf F}({\bf u})={\bf L}({\bf u})+{\bf N}({\bf u},{\bf u}).
\label{eq_fdef}
\end{equation}

In what follows, all that we will explicitly require of ${\bf F}$ is the 
linearity of ${\bf L}$ and the
fact that ${\bf N}$ is quadratic and symmetric in its arguments.
The following considerations will thus be valid for any ${\bf L}$ satisfying
\begin{equation}
{\bf L}(\lambda {\bf u_1}+\mu {\bf u_2})=\lambda {\bf L}({\bf u_1})+\mu {\bf L}({\bf u_2})
\label{eq_Llin}
\end{equation}
and quadratic ${\bf N}$ satisfying  
\begin{equation}{\bf N}({\bf u_1},{\bf u_2})={\bf N}({\bf u_2},{\bf u_1}).
\label{eq_Nsym}
\end{equation}
The main applications we have in mind are high order spatial discretizations
of the Navier-Stokes equation, the incompressible Euler equation 
\eq{eq_discrt}, the magnetohydrodynamics (MHD) equations, or
similarly nonlinear systems. Thus, the (real) vector ${\bf u}$ represents all of the (complex) components (modes) ${\hat v}_\alpha({\bf k})$ in a pseudo--spectral treatment, or 
all nodal or modal values in a spectral element or other high--order 
discretization. In these cases
${\bf L}$ and ${\bf N}$ can be obtained readily from the relevant terms in the
discrete systems. Naturally, the same schemes may be useful in low order or fixed
truncation methods as well.

\subsection{JST loop}
\label{ssec:jstloop}
All of our new algorithms start by using the current value of the field
${\bf u}= {\bf u}(t)$, to compute the order-$s$ JST loop \cite{Canuto-Hussaini}
\begin{eqnarray}
&&{\bf u}^*\gets {\bf u}~ \nonumber \\
&&\ {\rm For \ } k=s,1,-1 ~  \nonumber \\
&&\ {\bf u}^* \gets {\bf u} + \Delta t  \frac{{\bf F}({\bf u}^*)} {k}~  \nonumber \\
&&\ {\rm End \ For}~ .
 \label{eq_JSTLOOP}
\end{eqnarray}

The original order-$s$ JST algorithm \cite{Canuto-Hussaini} simply amounts to setting, after
the JST loop \ref{eq_JSTLOOP},
\begin{equation}
 {\bf u}^{\rm JST}(t+\Delta t)= {\bf u}^* .
\label{eq_JST}
\end{equation}
In the special case of linear ${\bf F}$, where ${\bf N}=0$, this algorithm can be obtained directly by 
factorizing the Taylor expansion
\begin{equation}
 {\bf u}(t+\Delta t)= {\bf u}(t)+\sum_{k=1}^s (\Delta t) ^{k}\frac {{\bf u^{(k)}}(t)} {k!}+\mathcal{O}(\Delta t ^{s+1})
\label{eq_Taylor}
\end{equation}
into
\begin{equation}
 {\bf u}(t+\Delta t)= \prod_{k=1}^s (1+\frac {\Delta t } {k}\frac {d} {dt}){\bf u}(t)
\label{eq_Prod}
\end{equation}
and is therefore exact. 
However, it is easy to check by explicit computation that when ${\bf N}\ne 0$, 
errors are present, beginning at order $3$ for $s \ge 3$. 

Indeed, defining the error term by
\begin{equation}
\delta{\bf u}= {\bf u}^{\rm JST}(t+\Delta t)- {\bf u}^{\rm exact}(t+\Delta t) \ ,
\label{eq_err3_def}
\end{equation}
it is straightforward to obtain explicit Taylor expansions in time
for both $ {\bf u}^{\rm exact}(t+\Delta t)$ and
$ {\bf u}^{\rm JST}(t+\Delta t)$
respectively from the evolution equation \eq{eq_evol} and the 
definitions \eq{eq_JSTLOOP} and \eq{eq_JST}, and compute the difference \eq{eq_err3_def}. For example, using obvious index notation for the rank-$n$ vectors ${\bf u}$ and ${\bf F}$, we obtain for the $i$ component when $s\ge 3$
\begin{equation}
\delta  u_i =-\frac{ \Delta t^3}{24} \sum_{j,k=1}^n \frac {\partial F_i}{\partial u_j \partial u_k } F_j F_k +\mathcal{O}(\Delta t ^{4}) .
\label{eq_err3_res}
\end{equation}
Note that this expression suggests that the local error is $\mathcal{O}(\Delta t ^{3})$; in 
almost all problems of interest, we integrate to a finite time, so the global error will then
be $\mathcal{O}(\Delta t ^{2})$.
The basic idea of the new algorithms we propose below is to modify the JST loop in order to cancel 
the $\mathcal{O}(\Delta t ^{3})$ term (and higher order terms) in \eq{eq_err3_res} that arises in the
presence of nonlinear terms in the evolution equation.

\subsection{Correction terms}
In the following it will be convenient to vary the number of iterations of the JSP algorithm independently of the order of the desired algorithm. As a result, we will refer to calculations made with the original algorithm at arbitraty iteration count $s$ as a JST-$s$ scheme. Recall that for nonlinear problems, all these algorithms have second order global truncation errors.

Using \eq{eq_err3_res} we immediately arrive at a new $3^{\rm rd}$--order algorithm setting, after $s=3$ JST iterations, 
\begin{eqnarray}
&{\bf u}&\gets{\bf L}({\bf u})+{\bf N}({\bf u},{\bf u})~ \nonumber \\
&{\bf u}&\gets2{\bf N}({\bf u},{\bf u}) ~  \nonumber \\
&{\bf u}^*&\gets {\bf u}^*+\frac{ \Delta t^3}{24} {\bf u}~  \nonumber \\
&{\bf u}&\gets {\bf u}^*~ .
 \label{eq_JST3PLUS}
\end{eqnarray}

Empirically we find that by increasing the number of JST iterations at a 
given truncation order, the overall error of the result is reduced. 
This can be shown to be the result of partial cancellations in 
higher order tems in \eq{eq_err3_res}. 
For example, if we use $s=4$ JST iterations (a JST-4 algorithm), and 
then apply the recipe \eq{eq_JST3PLUS}, we reduce the errors still further, 
even though the scheme is still manifestly $3^{\rm rd}$ order. We denote 
these cases with a $+$, where the number of $+$ following the order indicate 
how many extra JST iterations were done. As a result, the $3^{\rm rd}$--order 
method with $s=4$ is refered to as $3+$ in the discussion below.

Following the same procedure, we can compute $4^{\rm th}$--order 
correction terms. Higher order terms in \eq{eq_err3_res} can be 
canceled by making use of reasonable extra computational resources. 
Thus, a new $4^{\rm th}$--order algorithm requires that after the JST-4 
algorithm we set
\begin{eqnarray}
&{\bf u}&\gets{\bf L}({\bf u})+{\bf N}({\bf u},{\bf u})~ \nonumber \\
&{\bf u}&\gets {\bf u}+\frac{ \Delta t}{2}[{\bf L}({\bf u})+2{\bf N}({\bf u^*},{\bf u})] ~  \nonumber \\
&{\bf u}&\gets2{\bf N}({\bf u},{\bf u}) ~  \nonumber \\
&{\bf u}^*&\gets {\bf u}^*+\frac{ \Delta t^3}{24}{\bf u}~  \nonumber \\
&{\bf u}^*&\gets {\bf u}^*+\frac{ \Delta t^4}{72}[{\bf L}({\bf u})+2{\bf N}({\bf u^*},{\bf u})]~  \nonumber \\
&{\bf u}&\gets {\bf u}^*~ .
 \label{eq_JST4}
\end{eqnarray}
Again, if \eq{eq_JSTLOOP} is executed for $5$ iterations before applying 
\eq{eq_JST4}, we generally see a reduction in the overall error; hence, this
approach is called the $4+$ scheme.

Table \ref{Table} contains a summary of some of the different possibilities. The rows correspond to the number $s$ of JST loop iterations and the columns to the order of the correction terms 
\eq{eq_JST3PLUS} or  \eq{eq_JST4}. The number of evaluations of nonlinear terms and the order of the resulting method are indicated. Extra $+$ symbols follow the previously defined convention and are also related to the amount of error present in the numerical examples (\cf  \ \sec{SEC:NumRes}).

\begin{table}[h]
\begin{tabular}{|l||l|l||l|l|llll|}
\hline
$s$&\multicolumn{2}{l|}{No correction}&\multicolumn{2}{l|}{Third order}&\multicolumn{2}{l|}{Fourth order}\\
\cline{2-7}
&Order&$n_{\rm NL}$&Order&$n_{\rm NL}$&Order&$n_{\rm NL}$\\
\hline\hline
2&$2^{\rm nd}$&2&*&*&*&*\\
3&$2^{\rm nd} +$&3&$3^{\rm rd}$&5&*&*\\
4&$2^{\rm nd} ++$&4&$3^{\rm rd} +$&6&$4^{\rm th}$&8\\
\hline 
\end{tabular}
 \caption{Number of evaluations of nonlinear terms $n_{\rm NL}$ and order of the method obtained with $s$ 
JST iterations \eq{eq_JSTLOOP} using, respectively,  no correction,  $3^{\rm rd}$ order \eq{eq_JST3PLUS}, and $4^{\rm th}$ order \eq{eq_JST4} correction terms.}
 \label{Table}
 \end{table}

\section{Numerical results} \label{SEC:NumRes}
We now test these new algorithms on both conservative and dissipative systems.
The conservative systems are the $2$--degree--of--freedom classical
mechanics H\'enon--Heiles system 
and the full $3$-D spatially-periodic incompressible Euler equation.
Two dissipative systems are described by the Burgers equation.
The first dissipative example uses a standard 1-$D$ Fourier pseudo-spectral method, while the second a 2-$D$ spectral element method.

\subsection {Conservative systems}

\subsubsection {H\'enon-Heiles}
\label{sec:hheiles}
The H\'enon--Heiles Hamiltonian was introduced in 1964 \cite{Henon-Heiles} 
as a mathematical model of the chaotic motion of stars in a galaxy.
It is one of the simplest Hamiltonian systems to display soft chaos in classical mechanics.
The H\'enon-Heiles  Hamiltonian reads
\begin{equation}
E=\frac{\dot{x}^2 + \dot{y}^2}{2} +\frac{x^2 + y^2 + 2x^2 y -\frac{2}{3} y^3 }{2} .
\label{eq_H_HenonHeiles}
\end{equation}
The associated nonlinear nonintegrable canonical equations of motion 
that, of course, exactly conserve the total energy $E$ are
\begin{eqnarray}
\ddot{x}&=&  -x - 2 x y  ~,  \nonumber \\
\ddot{y}&=&  - y  -x^2 + y^2   \ . \label{Eq:Henon}
\end{eqnarray}
These equations are of the general quadratic form \eq{eq_fdef}, so  the 
H\'enon-Heiles system \eq{Eq:Henon}
is thus perhaps one of the simplest non-trivial dynamical system in which to test our new algorithms.

\begin{figure}[ht!]
\begin{center}
\hspace{-0.cm} \includegraphics[width=9.5cm,height=6.cm,angle=0]{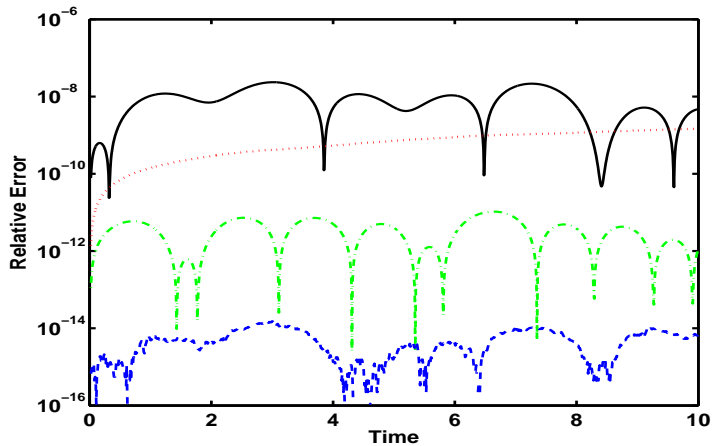}
\vspace{-0cm} 
\caption{\small{ Relative error in energy versus time in the H\'enon-Heiles system. Since the system is
conservative, the error should be zero.
Equations \eq{Eq:Henon} are solved and the resulting errors \eq{eq_H_HenonHeiles}
are shown corresponding to $2^{\rm nd}$--order ($3$ JST iterations; solid), $3^{\rm rd}$--order ($3$ JST iterations with 
correction; dotted), $3+$ ($4$ JST iterations with $3^{\rm rd}$--order correction; dash-dotted),
and $4^{\rm th}$--order ($4$ JST iterations with $4^{\rm th}$--order corrections; dash).}
\label{tHeno}}
\end{center}
\end{figure}

Figure \ref{tHeno} displays the numerical time-evolution of the relative error in the conserved energy 
$E(t)$ \eq{eq_H_HenonHeiles}, with $\Delta t=.001$ and initial data
$x(0)=0$, $y(0)= 0.12$, $\dot{x} (0)=0.486239$, $\dot{y} (0)=0.018$ corresponding to $E_0=.125$ for all
cases. The error is computed as $(E(t)-E_0)/E_0$.

It is apparent in the figure that the level of error decreases as the order of the method is increased. 
We note further that secular errors that are present in the third order method are canceled in the $3+$ case.
As expected, the fourth order conservation is much more precise than the $3+$, in this case, at or below 
machine round--off.

\subsubsection {Three-dimensional incompressible Euler equations}
\label{sec:euler}
The (unit density) three-dimensional (3D) 
incompressible Euler equations,
\begin{eqnarray}
{\partial_t {\bf v}}  + ({\bf v} \cdot \nabla) {\bf v}& =&- \nabla p  ~,  \nonumber \\
 \nabla  \cdot {\bf v} &=&0 ~,
 \label{eq_euler}
\end{eqnarray}
obeyed by a spatially $2 \pi$-periodic velocity field
can be approximated \cite{Canuto-Hussaini} 
by a (large) number of ordinary differential equations (ODEs)
by performing a Galerkin truncation (${\bf \hat v}({\bf k})=0$ for 
$|{\bf k}| \leq k_{\rm max}$) on 
the Fourier transform  ${\bf v}({\bf x},t)=\sum {\bf \hat v}({\bf k},t) e^{i {\bf k}\cdot {\bf x}}$.

One thus needs to solve the finite system of ODEs for the complex variables 
${\bf \hat v}({\bf k})$ (${\bf k}$ is a 3D vector of relative integers $(k_1,k_2,k_3)$
satisfying $|{\bf k}| \leq k_{\rm max}$)
\begin{equation}
{\partial_t { \hat v}_\alpha({\bf k},t)}  =  -\frac{i} {2} {\mathcal P}_{\alpha \beta \gamma}({\bf k}) 
\sum_{\bf p} {\hat v}_\beta({\bf p},t) {\hat v}_\gamma({\bf k-p},t)\, ,
\label{eq_discrt}
\end{equation}
where ${\mathcal P}_{\alpha \beta \gamma}=k_\beta P_{\alpha \gamma}+k_\gamma P_{\alpha \beta}$ with 
$P_{\alpha \beta}=\delta_{\alpha \beta}-k_\alpha k_\beta/k^2$ and the convolution in \eq{eq_discrt} 
is truncated to $|{\bf k}| \leq k_{\rm max}$, $|{\bf p}| \leq k_{\rm max}$ and $|{\bf k}-{\bf p} | \leq k_{\rm max}$.
This time-reversible system exactly conserves the kinetic energy
$E=\sum_{k}E(k,t)$ and helicity $H=\sum_{k}H(k,t)$ where the energy
and helicity spectra, $E(k,t)$ and $H(k,t)$, are defined by
respectively averaging
 ${\frac1 2}|{\bf \hat u}({\bf k'},t)|^2 \,$ and ${\bf
\hat u}({\bf k'},t)\cdot{\bf \hat \omega}({\bf -k'},t)$ (with ${\bf
\omega=\nabla\times {\bf u}}$ the vorticity) on spherical shells of width $\Delta k = 1$.

Numerical solutions of equation \eq{eq_discrt} are efficiently
computed using a pseudo-spectral general-periodic code
\cite{PabloCode1,PabloCode2} with $64^3$ Fourier modes that is dealiased using
the standard $2/3$ rule \cite{Got-Ors} by spherical Galerkin truncation at
$k_{\rm max}=21$. The code is fully parallelized with
the message passing interface (MPI) library. 

\begin{figure}
\begin{center}
\includegraphics[width=6.5cm]{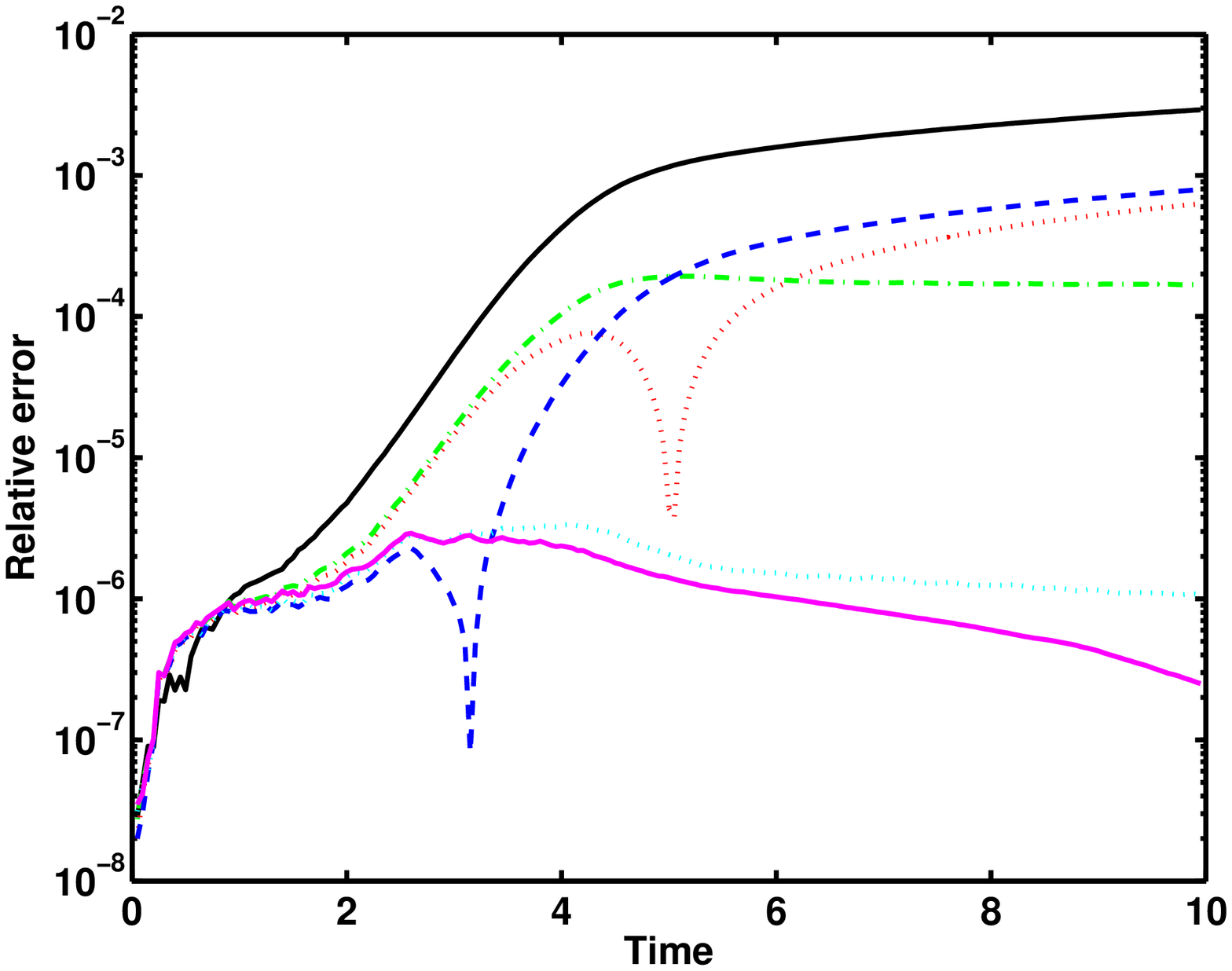}
\includegraphics[width=6.2cm]{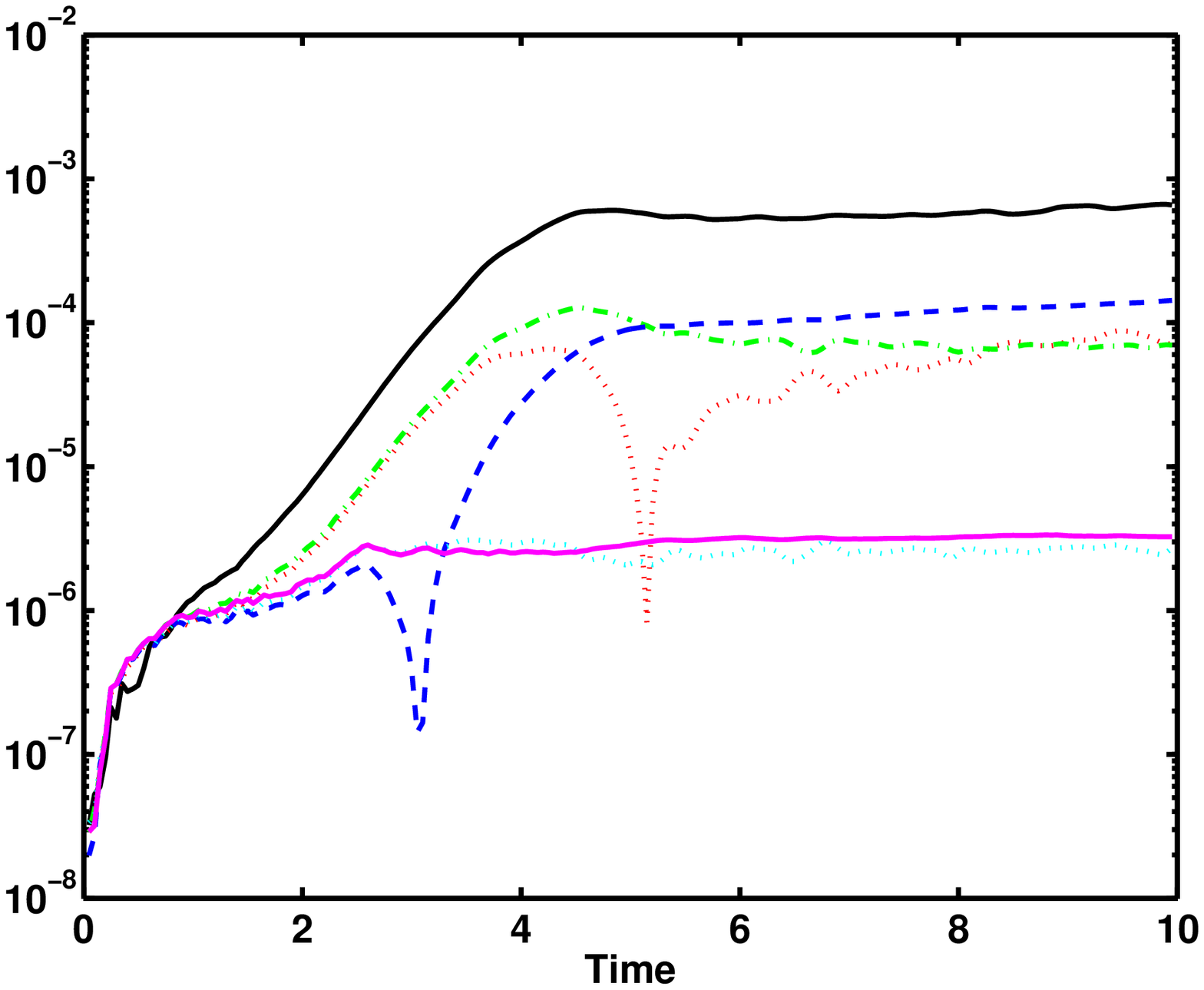}
\end{center}
\caption{\small{({\it left}) Energy conservation in terms of relative error versus time. The incompressible Euler 
equation is solved using a 
dealiased pseudospectral method with resolution $64^3$. Results are shown for JST-$2$, -$3$,
and -$4$ schemes (black, red dotted, and green dash-dotted curves, respectively),
$3^{\rm rd}$--order (blue dashed), $3+$ (cyan dotted), and $4^{\rm th}$--order
(solid magenta). ({\it right}) Helicity conservation in terms of relative error versus time, with curves representing
the same schemes as for energy conservation.}}
\label{tGdHd}
\end{figure}

The truncated Euler equation dynamics reaches,
by way of progressive thermalization \cite{CBDB-echel},
an absolute equilibrium that is a statistically stationary
gaussian exact solution of the associated Liouville equation
\cite{OrszagAnalytTheo}. \fig{tGdHd} displays 
the energy and helicity conservation during this process. As in the previous section, the error in the conservation is measured as the relative difference in the energy (helicity) compared to that at $t=0$, as in
\fig{tHeno}.

The first thing we notice is that the original JST-$s$ scheme (\ref{eq_JSTLOOP})
has secular errors that are clearly visible for $s=2$ and $s=3$ but generally not as pronounced 
for $s=4$ at late time ($t>6$). We also observe that both JST--$3$ and
the $3^{\rm rd}$--order schemes behave monotonically up to a certain time, then begin to increase its global error.
In fact, the global error of the $3^{\rm rd}$--order algorithm begins to exceed that of the JST-$2$ and JST-$3$ schemes at around $t=4$. Before $t \approx 2.5$, however, we see convergence of the errors that
behave with the new schemes roughly as they do in \fig{tHeno}.
As expected, we see the lowest errors when the method is increased to $4^{\rm th}$--order;
however, it is clear that part of the errors present in the pure $3^{\rm rd}$--order scheme are
canceled in the  $3+$ case. Note that the $3+$ and $4^{\rm th}$--order schemes yield errors
that decrease with time for this problem, a feature which may become important for very long integration
times.

\subsection {Dissipative systems}

\subsubsection{$1$D Burgers equation with a pseudospectral calculation}

\begin{figure}
\begin{center}
\includegraphics[width=6.5cm]{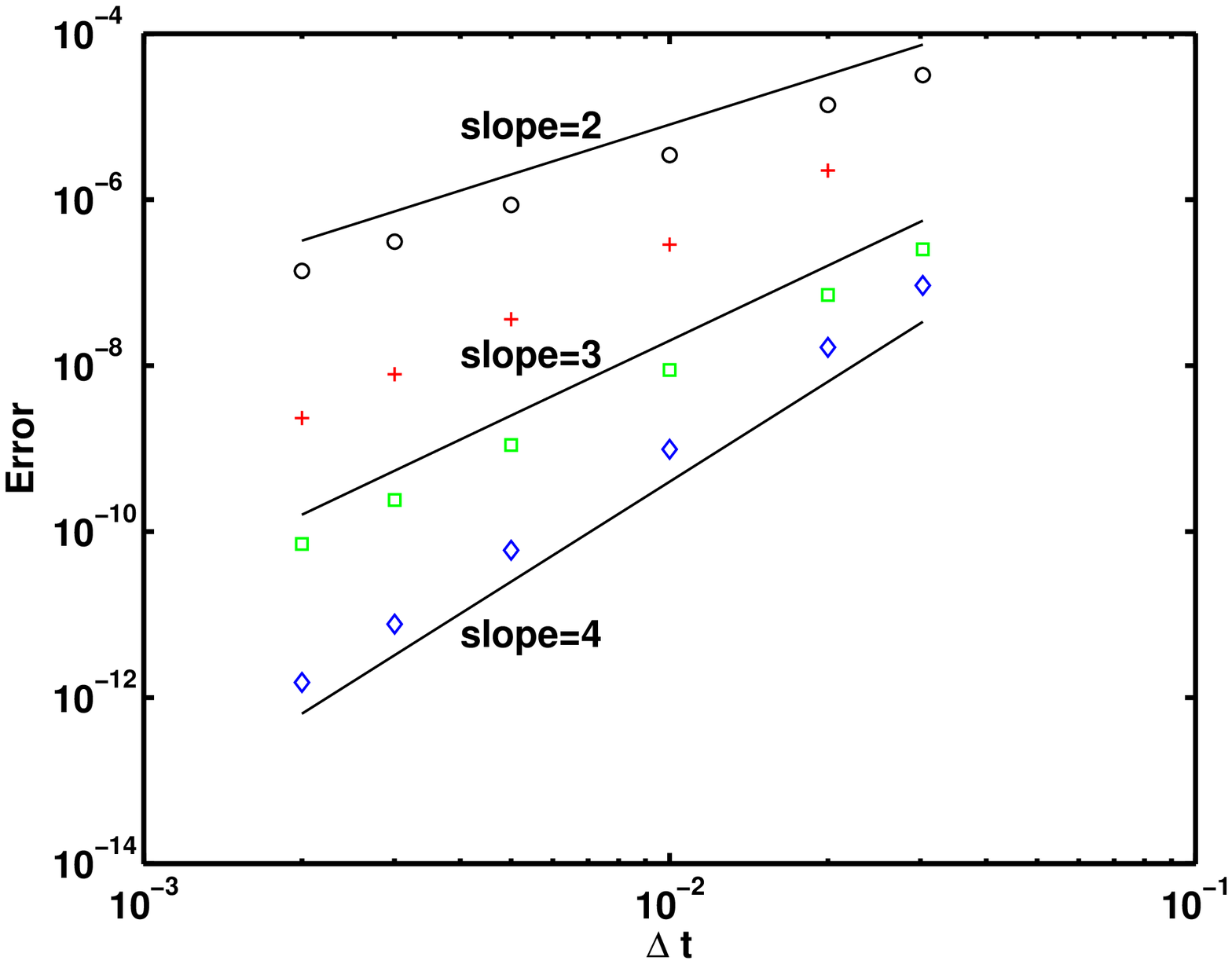}
\includegraphics[width=6.0cm]{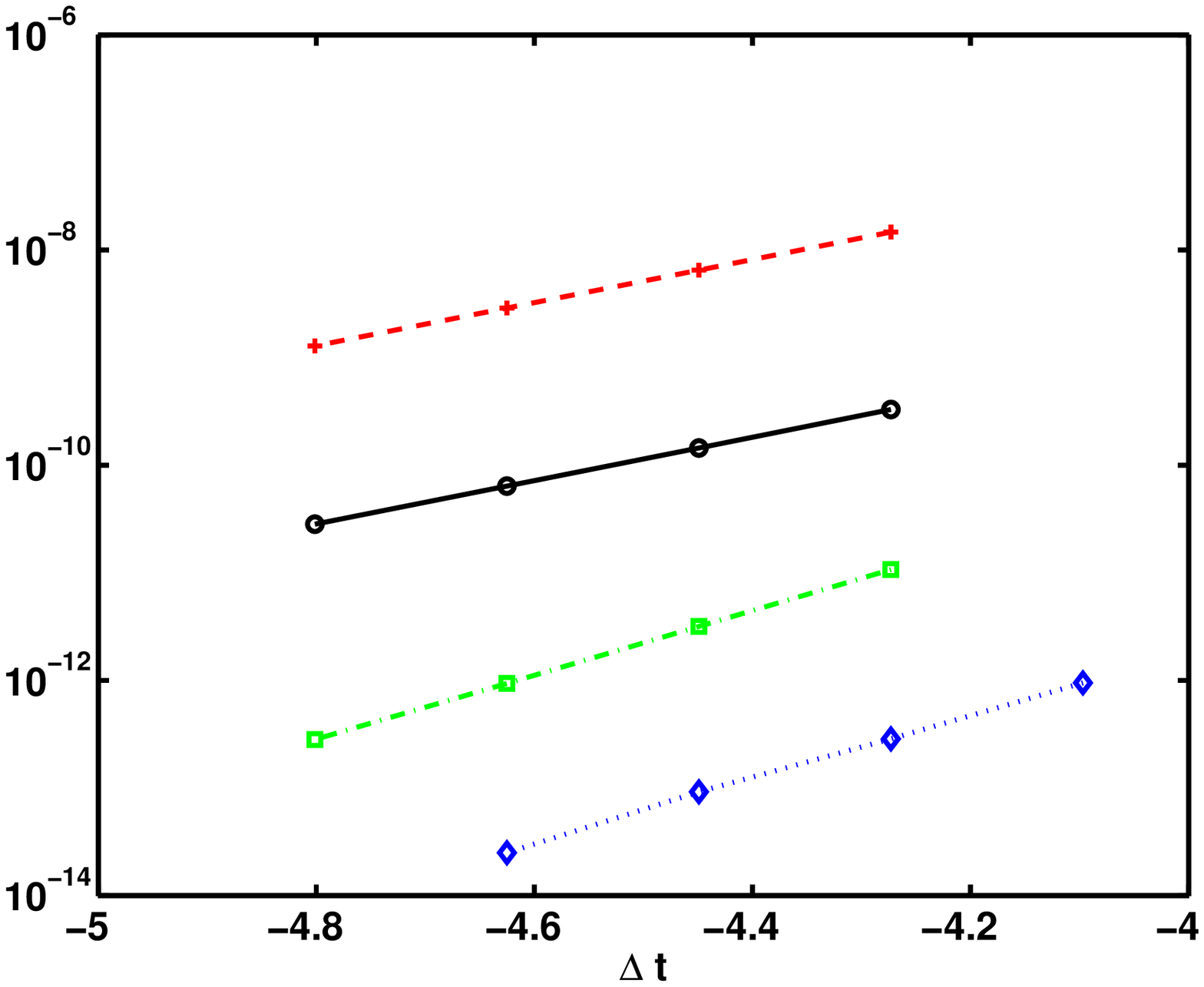}
\end{center}
\caption{({\it left}) \small{Absolute error of the slope of the Burgers front $\sup_x (-\partial_x v)$ 
at $t=2$ versus time-step $\Delta t$. Burgers equation with initial data $\sin(x)$ is solved using a 
dealiased pseudospectral method with resolution $N=64$, $\nu=2 \pi /N$. Results are shown for JST-$4$, 
(no corrections; circles), third order corrections (crosses), $3+$ (squares), and fourth order corrections 
(diamonds). Straight lines show the slopes indicating convergence orders 2, 3 and 4.}
({\it right}) \small{Absolute error in the area under the 1d curve (from \cite{whitham74}) vs time
step for a spectral element N-wave solution to Burgers equation.
Top red curve (slope $2.0017$): JST-$2$ algorithm; black curve:  JST-$3$
(slope $2.0181$); green curve: $3^{\rm rd}$--order 
(slope $2.989$). The blue curve (bottom) uses the $3+$ algorithm (slope $2.9716$).
}
}
\label{tPdSe}
\end{figure}

The $1$D Burgers equation
\begin{equation}
\partial_t u(x,t)+u(x,t)\partial_x u(x,t)=\nu \partial_{xx} u(x,t) , 
\label{eq_burgers}
\end{equation}
is of the general form (\ref{eq_fdef}).
It is solved here using a standard pseudo-spectral code
with $64$ Fourier modes, dealiased using
the $2/3$ rule \cite{Got-Ors}. We run to $t=2$ by which time a sharp front has formed for the chosen initial conditions $u(x,t=0)=\sin x$, a front whose width is
related to the viscosity, $\nu$. Each run is made with a different 
time step $\Delta t$, in order to check the truncation error as a function of $\Delta t$. We use as an error measure
the absolute difference between the slope of the front in the numerical and analytical (see, \eg, \cite{whitham74}) solutions.  

In \fig{tPdSe} we present the Burgers front truncation errors. We see immediately that the
errors decrease generally with an increase in the order of the scheme. In addition it is clear that here,
as in \fig{tGdHd}, the $3+$ scheme, while still $3^{\rm rd}$--order, can produce global errors that are
significantly lower than the $3^{\rm rd}$--order scheme alone.

\subsubsection{$1$D Spectral elements}

The spectral element method \cite{patera84} combines the flexibility of finite elements with the spectral convergence
of the pseudo--spectral method. Functions are expanded in each element in terms of Lagrange interpolating polynomials
(here the Gauss--Lobatto--Legendre polynomials), and $\Czero$  continuity conditions are imposed on the element
interfaces so that the functions reside in $\Hone$. 
The implementation we use is described in \cite{rosenberg2006}, and draws heavily from works
by other investigators \cite{mavriplis93,henderson99,fischer_kruse_2002,hfeng2002,DFM2002} in order to develop 
a new dynamic {\it h}--adaptive mesh refinement (AMR) formulation whereby the elements are
subdivided isotropically according to a variety of a-posteriori refinement (and coarsening) conditions.
The implementation sets the same polynomial degree in all elements, although this is not required of the method, 
and the polynomial degree can be varied for each run. The code forms a framework for solving a variety of
PDEs.
In addition, it is scalable, and is parallelized also by way of MPI. 
For this work, we use the nonlinear advection--diffusion solver, which solves the multi--dimensional
Burgers equation. The solver allows the use of semi--implicit and
explicit time stepping methods; an existing $2^{\rm nd}$--order JST method was modified to include 
trivially the third order correction terms \eq{eq_JST3PLUS}, with no additional storage. 

For this test, we solve the N-wave problem \cite{whitham74} on a 2D mesh without adaptivity. 
The 2D Cole-Hopf transformation
\begin{equation}
u=-2\nu\grad \ln \chi
\label{eq_colehopf}
\end{equation}
transforms \eq{eq_burgers} into a heat equation for $\chi$.
Choosing a source solution \cite{whitham74}
$$
\chi(x,t)=1+\frac{a}{t}\exp-\frac{(x-x_{\,0})^2}{4\nu t},
$$
we obtain the solution to \eq{eq_burgers} immediately from \eq{eq_colehopf}:
\begin{equation}
u(x,t)=\frac{x-x_{\,0}}{t}\frac{a}{a+t\exp((x-x_{\,0})^2/4\nu t)}.
\label{eq_nwave}
\end{equation}
This N-wave solution is essentially a 1D solution, but solved using a 2D solver with and without
the new high order time integration schemes. We choose $\nu=0.1$, $a=0.01$, and initial time $t=0.04$, and
integrate \eq{eq_burgers} to $t=0.06$ using a polynomials of degree $8$ in 
each direction.
In \fig{tPdSe} we consider the absolute error in the area under the surface \cite{whitham74} vs time step
in order to demonstrate the temporal convergence orders of the schemes.

\section{Conclusion} \label{SEC:Conc}

We have shown that for quadratically nonlinear equations of motion
the JST algorithm  \cite {JST}, \cite{Canuto-Hussaini} needs  corrections 
to go beyond second order.
We have computed the correction terms that enable the JST algorithm to be
used for $3^{\rm rd}$- and $4^{\rm th}$-order truncation errors in non-linear quadratic equations, and we show
that by utilizing the original JST algorithm, the new algorithm up to 
$4^{\rm th}$-order can be implemented with low storage requirements.

Numerical solutions to conservative and dissipative systems were used to 
verify the truncation errors of the new schemes, to verify their
stability properties, and to demonstrate that they
may be implemented easily using existing RK schemes. We considered implementations
for a $3$D pseudo--spectral incompressible Euler code, a $1$D pseudo--spectral
Burgers code, and a $2$D spectral element code.
We find that the most cost effective method is the $3+$ scheme, which requires
$4$ JST iterations and includes third order correction terms as described in the
text.

A natural question to ask is whether such high order explicit integration schemes 
are required. We hinted at an answer in discussing our 3D results: for
problems which are highly resolved spatially, as in pseudo--spectral or other
spectrally convergent discretizations, the time error can come to dominate the 
dynamics. This is clearly the case in \fig{tGdHd}, which shows that 
a low order time integration scheme integrated for a long time
could produce spurious conservation properties, yielding an unphysical result.

{\bf{Acknowledgments}}

The National Center for Atmospheric Research is sponsored by the National
Science Foundation.


\bibliographystyle{unsrt}
\bibliography{bibli}







\end{document}